\documentclass{elsart}

\usepackage{graphicx}
\usepackage{amssymb}

\newif\ifpreprint
\preprinttrue

\begin{document}

\journal{Physics Letter A}

\begin{frontmatter}

\date{November 6, 2009}
\title{
\ifpreprint
\fi
Measurement of the thorium-228 activity in solutions cavitated by ultrasonic sound}


\author{R.~Ford},
\author{M.~Gerbier-Violleau\thanksref{tra}},
\author{E.~V\'azquez-J\'auregui\corauthref{cor}}
\corauth[cor]{Corresponding author:}
\ead{ericvj@snolab.ca}

\address{SNOLAB, 1039 RR-24, Lively ON, P3Y 1N2, Canada}
\thanks[tra]{Present address: IPB, Institut Polytechnique Bordeaux, France}

\begin{abstract}
We show that cavitation of a solution of thorium-228 in water does not induce its transformation at a faster rate than the natural radioactive decay. We measured the activity of a thorium-228 solution in water before, and after, it was subjected to a cavitation at $44$ kHz and $250\,$W for $90$ minutes in order to observe any change in the thorium half-life. The results were compared to the original activity of the sample and we observed no change. Our results and conclusions conflict with those in a recent paper by F. Cardone et. al.~\cite{piezonuclear:2009pi}.  
\end{abstract}

\begin{keyword}
\PACS
23.90.+w
\sep
43.35.Ei
\end{keyword}

\end{frontmatter}

\section{Introduction}

A group of researchers at Rome, Italy, have investigated variations in nuclear decay and reaction rates induced by extreme pressure~\cite{piezonuclear:2009pi,piezo2:cardone2,piezo3:cardone3}. Such a revolutionary discovery would have an incommensurable impact in nuclear physics and its applications. A recent publication by F. Cardone et. al.~\cite{piezonuclear:2009pi} states that the rate of natural radioactive decay of thorium-228 can be increased up to a rate $10^4$ times faster when a solution of thorium in water is cavitated for $90$ minutes at a frequency of $20$ kHz and 100W. We designed an experiment to cavitate a solution of thorium-228 in water, and then mix with liquid scintillator to allow measurement of the radioactivity with a photomultiplier tube. We used a high sensitivity $\beta - \alpha$ particle coincidence counter and the technique of pulse shape discrimination to compare the activity before and after cavitation, and to confirm unbroken secular equilibrium in the thorium chain.

\section{Experiment}

We prepared five different "spike" solutions of thorium-228 mixed with Ultra Pure Water (UPW) and Optiphase Hi-Safe 3 liquid scintillator. The solutions were sealed in transparent plastic counting pots, which are 60ml hard plastic polymethylpentene jars with the bottom machined flat for efficient coupling to a photomultiplier tube face. The thorium-228 sources were obtained from several older sources we had on hand with residual activities ranging from 1-20 Bq/ml in 0.1M nitric acid (HN0$_3$). Each solution was made with $1\,$ml of thorium-228 source, and $10\,$ml of UPW, and then the remainder of the pot filled with scintillator, which was about 42g. This fraction, of up to 20\% aqueous solution added to the scintillator, is known to give good and reproducible light output efficiency~\cite{sno1:htio}. The filled pots have almost no air gap, thereby minimizing radon gas ingress, and the screw-top lids are sealed tight with para-film plastic.\\

Three of the solutions were mixed with UPW and scintillator at the start, and the other two were mixed only with UPW, with the scintillator added after cavitation. Additionally, four pots with only UPW and scintillator were prepared and labeled as blank samples.\\ 

We used a custom $\beta - \alpha$ particle coincidence counter (developed for the Sudbury Neutrino Observatory, see~\cite{sno1:htio}) to measure the activity of the thorium before and after cavitation. The device measures coincidences between $\beta$ and $\alpha$ particles from decay daughters in the Th-232 and U-238 chains. In the thorium chain Bi-212 decays to a short-lived state of Po-212 by $\beta$-particle emission with a branching ratio of $0.64$, which then decays to Pb-208 by $\alpha$-particle emission with $0.3$ $\mu s$ half-life (see Figure~\ref{th228}). In the Uranium chain Bi-214 decays to Po-214 by $\beta$-particle emission, followed by decay to Pb-210 by $\alpha$-particle emission with a $164$ $\mu s$ half-life. The thorium chain activity of a solution was measured by mixing with liquid scintillator, sealing in the transparent counting pot, and then coupling with vacuum grease to a photomultiplier tube (an Electron Tubes Ltd 9266XB). The $\alpha$-particles have a higher loss of energy per unit path length in the scintillator than the $\beta$-particles, and as a result the $\alpha$'s have a longer pulse decay time than the $\beta$'s. It can be observed that a larger fraction of the total charge of the pulse will be in the tail for the $\alpha$-particle, and if it is compared to the $\beta$-particle this can be used to discriminate between them by Pulse Shape Discrimination (PSD). The counter records an event based on the PSD identification of a $\beta$-particle followed by an $\alpha$-particle within a several $\mu$sec coincidence window. This coincidence and PSD technique provides almost complete elimination of backgrounds due to $\beta - \beta$ events, $\beta - \gamma$ events, cosmic rays, coincident backgrounds, and partial tracks. The custom electronics measures the coincidence time, so that thorium chain events are distinguished from uranium chain events, and eliminates the issue of contamination from Rn-222. The sensitivity of the counters is about 25 mBq. Disequilibrium in the chains can also be measured by counting over a period of a week or more, then plotting the activity as a function of time (the time histogram), and fitting with the Ra-224 ($t_{1/2} = 3.6\,$d) and Pb-212 ($t_{1/2} = 10.6\,$h) decay functions. A complete description of the counter system and PSD analysis can be found elsewhere~\cite{sno1:htio}.\\

The cavitation device used for the experiment was a Fisher Scientific FS220H ultrasonic cleaner with an operating frequency of $44$ kHz and an power of $250\,$W. The process was to fill and seal the pots and place in the Ultrasonic bath within the steel perforated basket and the bath filled with UPW. The three thorium solutions with the scintillator were cavitated for $90$ minutes and then counted for $15$ minutes, and then the cavitation and counting was repeated a second time. One of the thorium solutions with only UPW was cavitated for $90$ minutes once only, and then scintillator was added to the pot in order to count the activity. The blank samples (the pots with only scintillator and water) were also cavitated twice for $90$ minutes. During the cavitation process the temperature of the water in the ultrasonic bath increased from $20^o$C to $40^o$C. Note that the $\beta - \alpha$ particle counter electronics are multi-channel, and we used three channels to count three pots simultaneously.

\section{Data Analysis}

\begin{figure}[hbtp]
\centering
\includegraphics[angle=0,width=0.60\textwidth]{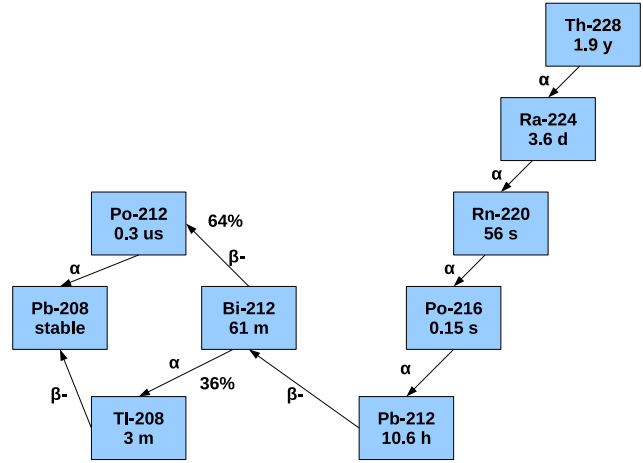}
  \caption{Thorium-228 decay chain.}
  \label{th228}
\end{figure}

The decay chain of thorium-228 can be depicted as shown in Figure~\ref{th228}, where the $\beta - \alpha$ particle counter records the coincidence events with the $\beta$ decay of Bismuth-212 and the $\alpha$ decay of the Polonium-212 (called Bi-Po events). The thorium chain Bi-Po activity of the spiked solutions was measured before and after cavitation, with the results summarized in Table~\ref{tab:spikes}. Spike $5$ was not counted before cavitation, as scintillator would be needed, and we wanted to cavitate a solution with exclusively water in it. So in order to compare the activity for spike $5$ before and after cavitation, spike $4$ was used as reference (counted before) since samples $4$ and $5$ were prepared with the same activity.\\

\begin{table}[htb]
\centering
\vspace{0.5cm}
\begin{tabular}{|c|c|c|l|}
\hline\hline
Sample & Before (cps) & After (cps) & Cavitation Time\\
\hline\hline
spike $1$ & $16.4 \pm 1.8$ & $16.5 \pm 1.9$ & 90 min\\
             &  & $16.5 \pm 1.9$ & 90 min (second time)\\
\hline
spike $2$ & $1.2 \pm 0.2$ & $1.2 \pm 0.2$ & 90 min\\
             &  & $1.2 \pm 0.2$ & 90 min (second time)\\
\hline
spike $3$ & $1.7 \pm 0.2$ & $1.7 \pm 0.2$ & 90 min\\
             &  & $1.5 \pm 0.2$ & 90 min (second time)\\
\hline
spike $4$ & $2.6 \pm 0.3$ & & Not cavitated\\
 & & & (reference for $5$)\\
spike $5$ &  & $2.5 \pm 0.3$ & 90 min\\
\hline
\end{tabular}
\vspace{0.5cm}
\caption{Activity of the spiked samples before and after cavitation. Samples 1, 2 and 3 were solutions of Th-228 in water and scintillator cavitated twice for $90\,$ minutes. Sample 5 was the solution of Th-228 in water cavitated once for $90\,$ minutes. Sample 4 was not cavitated; it was labeled as reference for sample 5, since they were prepared under the same conditions. Scintillator was added to samples 4 and 5 just before counting. Counting time was 15 minutes, and the uncertainties are statistical combined with uncertainty in the counter efficiency.}
\label{tab:spikes}
\end{table}

A comparison of the activities before and after the cavitation for the blank samples is shown in Table~\ref{tab:blanks}. The samples were each counted for two days. The results show that the thorium-228 activity in the scintillator is small compared to the activity of the spiked samples, and that the value remains constant after $90\,$ minutes in the ultrasonic bath. This simply shows that the scintillator has low background and that the properties of the scintillator are not affected by the cavitation. It is important to clarify that the total number of counts are reported for the blank samples after two days in the $\beta - \alpha$ particle counter, while for the spike solutions the counts per second are reported.\\

\begin{table}[htb]
\centering
\vspace{0.5cm}
\begin{tabular}{|c|c|c|c|}
\hline\hline
Sample & Before (counts) & After (counts)& Cavitation Time\\
\hline\hline
blank $1$ & $21 \pm 5$ & $23 \pm 5$ & 90 min\\
blank $2$ & $11 \pm 3$ & $6 \pm 2$ & 90 min\\
blank $3$ & $11 \pm 3$ & $10 \pm 3$ & 90 min\\
blank $4$ & $10 \pm 3$ & $3 \pm 2$ & 90 min\\
\hline
\end{tabular}
\vspace{0.5cm}
\caption{Activity of the blank samples before and after cavitation, as total counts over two days. Each solution is approximately 10g UPW and 42g scintillator. The uncertainties are statistical combined with uncertainty in the counter efficiency.}
\label{tab:blanks}
\end{table}

\section{Discussion and Conclusions}

The results summarized in Table~\ref{tab:spikes} show that the counts per second for the Th-228 solutions are the same before and after cavitation. This strongly suggests that cavitation did not affect the activity of thorium-228. We also counted and recorded the time dependence of the activity of the samples over the weeks following, to check if equilibrium was broken during cavitation, with radium-224 being produced at a higher or lower rate than natural, before regaining equilibrium. We consider that the radium-224 decay rate may also have changed during the cavitation period. However, no time dependence was observed, and we concluded that the chain is still in equilibrium and supported by the natural decay of thorium-228.\\

We showed that the cavitation of a thorium-228 solution in Ultra Pure Water at a frequency of $44\,$kHz for $90$ minutes does not change the half-life of the thorium. A change in the half-life of the thorium-228 or any other radioactive element would be a revolutionary result and it would have an enormous impact not only in the nuclear physics field but also its applications. We expect that our experiment will provide some welcome data for the wider discussions held in the physics community~\cite{com1:sweden08045,com2:sweden08047,com3:swedenarxiv} about the piezonuclear reactions of radioactive elements. We note that our investigation used an ultrasonic frequency of $44\,$kHz, whereas the F. Cardone et. al. experiment was at $20\,$kHz, although our experiment used more power. Even if there is a frequency dependence, there would need to be an incredibly sharp resonance to have a $10^4$ rate affect at $20\,$kHz with no change at $44\,$kHz. None of the discussions in the works cited suggest such a sharp resonance effect, and the probability of such a resonance being at the single frequency tested by F. Cardone et al. is unlikely.\\

\section*{Acknowledgments}

This research was supported by: Natural Sciences and Engineering Research Council, Northern Ontario Heritage Fund, and the Canada Foundation for Innovation. We also thank the SNOLAB technical staff for their invaluable assistance, as well as Laurentian University for hosting the experiment and helping in the preparation of the radioactive solutions. We thank Vale Inco for hosting SNOLAB where the main ideas for this experiment arose.

\end{document}